----------------------------------------------------------------
\documentstyle[aps,12pt]{revtex}
\begin{document}

\title{Generalized Radial Equations in a Quantum $N$-Body Problem}

\author{Zhong-Qi Ma \thanks{Electronic
address:MAZQ@IHEP.AC.CN}, Bing Duan, and Xiao-Yan Gu}

\address{Institute of High Energy Physics, Beijing 100039,
The People's Republic of China}

%\date{}

\maketitle

\vspace{5mm}
\begin{abstract}
We demonstrate how to separate the rotational degrees of
freedom in a quantum $N$-body problem completely from the
internal ones. It is shown that any common eigenfunction of
the total orbital angular momentum ($\ell$) and the parity
in the system can be expanded with respect to $(2\ell+1)$
base-functions, where the coefficients are the functions
of the internal variables. We establish explicitly the
equations for those functions, called the generalized radial 
equations, which are $(2\ell+1)$ coupled partial differential 
equations containing only $(3N-6)$ internal variables.

\vspace{3mm}
\noindent
PACS number(s): 11.30.-j, 03.65.Ge, and 03.65.Fd

\end{abstract}
\maketitle

\vspace{5mm}
Symmetry is an important property of a physical system.
The symmetry of a quantum system can simplify its
Schr\"{o}dinger equation and remove some variables in the 
equation. The simplest example is the hydrogen atom problem, 
where, due to the spherical symmetry, the wavefunction is 
expressed as a product of a radial function and a spherical 
harmonic function,
$$\Psi^{\ell}_{m}({\bf r})=\phi(r)Y^{\ell}_{m}(\theta,\varphi), \eqno (1)$$

\noindent 
and the Schr\"{o}dinger equation reduces to a radial
equation with only one radial variable. For a quantum $N$-body
problem with a pair potential, the Schr\"{o}dinger equation is
invariant under the spatial translation, rotation, and inversion.
From those symmetries, one should be able to separate the motion
of center-of-mass and the global rotation of the system from the
internal motions so as to reduce the Schr\"{o}dinger equation to
the generalized "radial" equation that contains only internal
variables. However, this problem has not been solved. In this
letter we will solve this problem completely. Using the
appropriately chosen $(3N-6)$ internal variables and the
$(2\ell+1)$ base-functions for the total orbital angular momentum
$\ell$, we establish explicitly the generalized radial equations
without any approximation. Only $(3N-6)$ internal variables are
involved in both the generalized radial functions and the
equations.

Denote by ${\bf r_{j}}$ the position vectors of $N$-particles 
with masses $m_{j}$ in the laboratory frame (LF), respectively.
The Schr\"{o}dinger equation for the $N$-body problem is
$$-\left(\hbar^{2}/2\right) \displaystyle
\sum_{j=1}^{N}~m_{j}^{-1} \bigtriangleup_{{\bf r}_{j}}\Psi +V \Psi
=E \Psi , \eqno (2) $$

\noindent 
where $V$ is assumed to be a pair potential, depending on the 
distances of each pair of particles. Therefore, the potential 
$V$ is a function of only the internal variables. It is well 
known that, due to the translation symmetry of the system,
the motion of center-of-mass can be separated completely from
others by making use of the Jacobi coordinate vectors in the 
center-of-mass frame (CF) [1-3], 
$${\bf R}_{k}=\left(\displaystyle {m_{k}W_{k+1}\over W_{k}} 
\right)^{1/2} \left({\bf r}_{k}-\displaystyle \sum_{j=k+1}^{N}~ 
\displaystyle {m_{j}{\bf r}_{j} \over W_{k+1}}\right), ~~~~~
1\leq k \leq (N-1). \eqno (3) $$

\noindent
where $W_{j}=\sum_{t=j}^{N}~m_{t}$. In CF, the Laplace operator
and the total orbital angular momentum operator ${\bf L}$
can be directly expressed with respect to ${\bf R}_{k}$:
$$\bigtriangleup =\displaystyle \sum_{j=1}^{N}~\displaystyle
m_{j}^{-1} \bigtriangleup_{{\bf r}_{j}}
= \displaystyle \sum_{k=1}^{N-1}~\bigtriangleup_{{\bf R}_{k}}, $$
$${\bf L}=-i\hbar \displaystyle \sum_{j=1}^{N}~ {\bf r}_{j}
\times \bigtriangledown_{{\bf r}_{j}}
=-i\hbar \displaystyle \sum_{k=1}^{N-1}~ {\bf R}_{k}
\times \bigtriangledown_{{\bf R}_{k}} , \eqno (4) $$

\noindent 
The Laplace operator obviously has the symmetry of $O(3N-3)$ 
group with respect to $(3N-3)$ variables. The $O(3N-3)$ group 
contains a subgroup $SO(3)\times O(N-1)$, where $SO(3)$ is
the usual rotational group. The space inversion and the different
definitions for the Jacobi vectors in the so-called "Jacobi tree"
[4] can be obtained by $O(N-1)$ transformations. For the system of
identical particles, the permutation group among particles is also
a subgroup of $O(N-1)$ group.

Because of the spherical symmetry, the angular momentum is
conserved. The hydrogen atom problem is a typical quantum two-body
problem, where there is only one Jacobi coordinate vector, usually
called the relative position vector ${\bf r}$. For a quantum
$N$-body problem, equation (1) should be generalized in three
aspects. The first is how to define the internal variables, which
describe the internal motions completely. The second is how to
find the complete set of the independent base-functions with the
given angular momentum. The total wavefunction is expanded with
respect to the base-functions, where the coefficients are the
generalized radial functions which only depend on the internal
variables. The last is how to derive the generalized radial
equations that only contain $(3N-6)$ internal variables. As a
matter of fact, these three aspects are connected. The parity
should also be considered in the generalization. Due to the
spherical symmetry, one only needs to study the eigenfunctions 
of angular momentum with the largest eigenvalue of $L_{3}$
($m=\ell$), which are simply called the wavefunctions with the
angular momentum ${\ell}$ in this letter for simplicity. Their
partners with the smaller eigenvalues of $L_{3}$ can be 
calculated from them by the lowering operator $L_{-}$.

Denote by $R=R(\alpha,\beta,\gamma)$ a spatial rotation,
transforming CF to the body-fixed frame (BF), and by $\xi$ all 
the internal variables in a quantum $N$-body problem for simplicity.
Although Wigner did not separate the motion of center-of-mass by
the Jacobi vectors, he proved from the group theory that any
wavefunction with the angular momentum $\ell$ in the system can
be expressed as follows (see Eq. (19.6) in [5]):
$$\Psi^{\ell}_{\ell}(\alpha,\beta,\gamma,\xi)=\displaystyle
\sum_{q=-\ell}^{\ell}~ D^{\ell}_{\ell
q}(\alpha,\beta,\gamma)^{*}\psi_{q}(\xi), \eqno (5) $$

\noindent 
where we adopt the commonly used form of the $D$-function [6]. 
In Eq. (5) $D^{\ell}_{\ell q}(\alpha,\beta,\gamma)^{*}$ plays 
the role of the base-function with the angular momentum ${\ell}$, 
and $\psi_{q}(\xi)$ is the generalized radial function. What 
Wigner proved is that there are only $(2\ell+1)$ independent 
base-functions with the angular momentum ${\ell}$. Unfortunately, 
due to the singularity of the Euler angles, the generalized 
radial equations are very difficult to derive based on Eq. (5). 
Wigner did not discuss the generalized radial equations, and 
to our knowledge, those equations have not yet been established 
in the literature. It is obvious that the generalized radial 
equations are very easy to obtain for the $S$ wave [7]. However, 
it seems quite difficult to obtain even for $P$ wave in a 
three-body problem [8,9].

Recently, a coupled angular momentum basis was used to
prediagonalize the kinetic energy operator [10], where some
off-diagonal elements remain non-vanishing. In their calculation,
the function with a given angular momentum was combined from the
partial angular momentum states by the Clebsch-Gordan
coefficients. Since the partial angular momenta are not conserved,
one has to deal with, in principle, an infinite number of the
partial angular momentum states. This problem also occurs in the
hyperspherical harmonic function method and its improved ones
[2,4,11,12]. However, as Wigner proved, only $(2\ell+1)$ partial
angular momentum states are involved in constructing the
base-functions with the angular momentum $\ell$.

Arbitrarily choose two Jacobi coordinate vectors, say ${\bf R}_{1}$
and ${\bf R}_{2}$. Let ${\bf R}_{1}$ be parallel with the $Z$-axis
of BF, and ${\bf R}_{2}$ be located in the $XZ$ plane with a
non-negative $X$-component in BF. The rotational degrees of freedom
of the system are described by a rotation $R(\alpha, \beta, \gamma)$,
transforming CF to BF. Define $(3N-6)$ internal variables, which
should be invariant in the global rotation $R(\alpha, \beta, \gamma)$:
$$\xi_{j}={\bf R_{j}}\cdot {\bf R}_{1},~~~~~
\eta_{j}={\bf R_{j}}\cdot {\bf R}_{2},~~~~~
\zeta_{j}={\bf R_{j}}\cdot \left({\bf R}_{1} \wedge 
{\bf R}_{2} \right),~~~~~1\leq j \leq (N-1), \eqno (6) $$

\noindent 
where $\eta_{1}=\xi_{2}$ and $\zeta_{1}=\zeta_{2}=0$. It
is worth mentioning that $\xi_{j}$ and $\eta_{j}$ have even
parity, but $\zeta_{j}$ has odd parity. From them we have
$$\Omega_{j}=\left({\bf R}_{1} \wedge {\bf R}_{j}\right)\cdot
\left({\bf R}_{1}\wedge {\bf R}_{2}\right)
=\xi_{1}\eta_{j}-\xi_{2}\xi_{j},$$ 
$$\omega_{j}=\left({\bf R}_{2}
\wedge {\bf R}_{j}\right)\cdot \left({\bf R}_{1}\wedge {\bf
R}_{2}\right) =\xi_{2}\eta_{j}-\eta_{2}\xi_{j},$$ 
$${\bf R}_{j}\cdot {\bf R}_{k}=\Omega_{2}^{-1}
\left(\Omega_{j}\eta_{k}-\omega_{j}\xi_{k}+\zeta_{j}\zeta_{k}\right),
\eqno (7) $$

\noindent
where $\Omega_{1}=\omega_{2}=0$, and
$\Omega_{2}=-\omega_{1}=\left({\bf R}_{1}\wedge {\bf R}_{2}\right)^{2}$.

Recall that two Jacobi vectors ${\bf R}_{1}$ and ${\bf R}_{2}$
completely determine BF and three Euler angles. The base-functions
with the angular momentum ${\ell}$ should be combined from the
products of two spherical harmonic functions $Y^{q}_{m}({\bf R}_{1})$
and $Y^{p}_{m'}({\bf R}_{2})$ by the Clebsch-Gordan coefficients
$\langle q,m,p,m'|\ell,(m+m')\rangle$. Define [3,11]
$$Q_{q}^{\ell \tau}({\bf R}_{1},{\bf R}_{2})
=\displaystyle {(R_{11}+iR_{12})^{q-\tau}
(R_{21}+iR_{22})^{\ell-q} \over (q-\tau)!(\ell-q)!}
\left\{(R_{11}+iR_{12})R_{23}-R_{13}(R_{21}+iR_{22})\right\}^{\tau}, $$
$$\tau \leq q \leq \ell,~~~~~\tau=0,1.  \eqno (8) $$

\noindent 
where $R_{ja}$ is the $a$th component of the Jacobi
vector ${\bf R}_{j}$. $Q_{q}^{\ell \tau}({\bf R}_{1},{\bf R}_{2})$
is the common eigenfunction of ${\bf L}^{2}$, $L_{3}$,
$\bigtriangleup_{{\bf R}_{k}}$, and the parity with the
eigenvalues $\ell(\ell+1)$, $\ell$, $0$, and $(-1)^{\ell+\tau}$,
respectively. As a matter of fact, the following combination of
products of two spherical harmonic functions is proportional to
$Q_{q}^{\ell \tau}({\bf R}_{1},{\bf R}_{2})$ 
$$\displaystyle \sum_{m}~ \xi_{1}^{q/2}Y^{q}_{m}({\bf
R}_{1})\eta_{2}^{(\ell-q+\tau)/2} Y^{\ell-q+\tau}_{\ell-m}({\bf
R}_{2}) \langle q,m,(\ell-q+\tau),(\ell-m)|\ell,\ell\rangle
=CQ_{q}^{\ell \tau}({\bf R}_{1},{\bf R}_{2}), \eqno (9) $$

\noindent
where $C$ is a normalization factor. Now, we come to the theorem.

\vspace{3mm}
\noindent
{\bf Theorem}. Any function
$\Psi^{\ell \lambda}_{\ell}({\bf R}_{1}, \cdots, {\bf R}_{N-1})$
with the angular momentum ${\ell}$ and the parity
$(-1)^{\ell+\lambda}$ in a quantum $N$-body problem can be
expanded with respect to
$Q_{q}^{\ell \tau}({\bf R}_{1},{\bf R}_{2})$ with the
coefficients $\psi^{\ell \lambda}_{q \tau}(\xi,\eta,\zeta)$,
which depend on $(3N-6)$ internal variables:
$$\Psi^{\ell \lambda}_{\ell}({\bf R}_{1}, \cdots, {\bf R}_{N-1})
=\displaystyle \sum_{\tau=0}^{1} \displaystyle \sum_{q=\tau}^{\ell}~
\psi^{\ell \lambda}_{q \tau}(\xi,\eta,\zeta)
Q_{q}^{\ell \tau}({\bf R}_{1},{\bf R}_{2}), \eqno (10) $$
$$\psi^{\ell \lambda}_{q \tau}(\xi,\eta,\zeta)=
\psi^{\ell \lambda}_{q \tau}(\xi_{1},\cdots,\xi_{N-1},\eta_{2},
\cdots, \eta_{N-1},\zeta_{3},\cdots,\zeta_{N-1}), $$

\noindent 
where the parity of $\psi^{\ell \lambda}_{q \tau}(\xi,\eta,\zeta)$ 
is $(-1)^{\lambda-\tau}$.

Equation (5) coincides with Eq. (10), because either of the set of
$D^{\ell}_{\ell q}(\alpha,\beta,\gamma)^{*}$ and the set of
$Q_{q}^{\ell \tau}({\bf R}_{1},{\bf R}_{2})$ is a complete set of
base-functions of the angular momentum. However, equation (10) has
three important characteristics, which make it possible to derive
the generalized radial equations. The first is that the Euler
angles do not appear explicitly in the base-functions 
$Q_{q}^{\ell \tau}({\bf R}_{1},{\bf R}_{2})$. The second is the 
well chosen internal variables (6). The third is that the internal 
variables $\zeta_{j}$ have odd parity. It is due to the existence of
$\zeta_{j}$ that the base-functions 
$Q_{q}^{\ell 0}({\bf R}_{1},{\bf R}_{2})$ and 
$Q_{q}^{\ell 1}({\bf R}_{1},{\bf R}_{2})$ appear together in one 
total wavefunction. By comparison, all the internal variables in 
a quantum three-body problem have even parity ($\zeta_{j}=0$) so 
that in a total wavefunction with a given parity, only the 
base-functions with the same parity appear [5,12].

Now, substituting Eq. (10) into the Schr\"{o}dinger equation (2)
with the Laplace operator (4), we obtain the generalized radial
equations by a straightforward calculation: 
$$\bigtriangleup \psi^{\ell \lambda}_{q0} 
+4\left\{q\partial_{\xi_{1}}+(\ell-q)\partial_{\eta_{2}} \right\}
\psi^{\ell \lambda}_{q0} 
+2q\partial_{\xi_{2}} \psi^{\ell \lambda}_{(q-1)0} 
+2(\ell-q)\partial_{\xi_{2}} \psi^{\ell \lambda}_{(q+1)0} $$ 
$$+\displaystyle \sum_{j=3}^{N-1}~2\Omega^{-1}_{2}\left\{
\left[-\omega_{j}q\partial_{\xi_{j}}
+\Omega_{j}(\ell-q)\partial_{\eta_{j}}
+\eta_{2}\zeta_{j}q\partial_{\zeta_{j}}
+\xi_{1}\zeta_{j}(\ell-q) \partial_{\zeta_{j}}\right] 
\psi^{\ell \lambda}_{q0}  \right. $$
$$-q\left[\omega_{j}\partial_{\eta_{j}}
+\xi_{2}\zeta_{j}\partial_{\zeta_{j}}\right] \psi^{\ell \lambda}_{(q-1)0}
+(\ell-q)\left[\Omega_{j}\partial_{\xi_{j}}-\xi_{2}\zeta_{j}
\partial_{\zeta_{j}}\right] \psi^{\ell \lambda}_{(q+1)0} $$
$$-i\eta_{2}q(q-1)\left[\zeta_{j}\partial_{\eta_{j}}
-\Omega_{j}\partial_{\zeta_{j}}\right] \psi^{\ell \lambda}_{(q-1)1}$$
$$-iq\left[\eta_{2}\zeta_{j}q\partial_{\xi_{j}}
-\xi_{2}\zeta_{j}(2\ell-2q+1)\partial_{\eta_{j}}
+\eta_{2}\omega_{j}q\partial_{\zeta_{j}}
+\xi_{2}\Omega_{j}(2\ell-2q+1)\partial_{\zeta_{j}}
\right]  \psi^{\ell \lambda}_{q1}$$
$$+i(\ell-q)\left[\xi_{2}\zeta_{j}(2q+1)\partial_{\xi_{j}}
-\xi_{1}\zeta_{j}(\ell-q)\partial_{\eta_{j}}
+\xi_{2}\omega_{j}(2q+1)\partial_{\zeta_{j}}
+\xi_{1}\Omega_{j}(\ell-q)\partial_{\zeta_{j}}
\right] \psi^{\ell \lambda}_{(q+1)1}$$
$$\left.-i\xi_{1}(\ell-q)(\ell-q-1)\left[\zeta_{j}\partial_{\xi_{j}}
+\omega_{j}\partial_{\zeta_{j}}\right] \psi^{\ell \lambda}_{(q+2)1}\right\}
=-\left(2/\hbar^{2}\right)\left[E-V\right]\psi^{\ell \lambda}_{q0},
\eqno (11a) $$
$$\bigtriangleup \psi^{\ell \lambda}_{q1}
+4\left\{q\partial_{\xi_{1}}+(\ell-q+1)\partial_{\eta_{2}}
\right\} \psi^{\ell \lambda}_{q1}
+2(q-1)\partial_{\xi_{2}} \psi^{\ell \lambda}_{(q-1)1}
+2(\ell-q)\partial_{\xi_{2}} \psi^{\ell \lambda}_{(q+1)1}$$
$$+\displaystyle \sum_{j=3}^{N-1}~2\Omega^{-1}_{2}\left\{
\left[-\omega_{j}q\partial_{\xi_{j}}+\Omega_{j}(\ell-q+1)\partial_{\eta_{j}}
+\eta_{2}\zeta_{j}q\partial_{\zeta_{j}}+\xi_{1}\zeta_{j}(\ell-q+1)
\partial_{\zeta_{j}}\right] \psi^{\ell \lambda}_{q1}\right. $$
$$-(q-1)\left[\omega_{j}\partial_{\eta_{j}}
+\xi_{2}\zeta_{j}\partial_{\zeta_{j}}
\right] \psi^{\ell \lambda}_{(q-1)1}
+(\ell-q)\left[\Omega_{j}\partial_{\xi_{j}}-\xi_{2}\zeta_{j}
\partial_{\zeta_{j}}\right] \psi^{\ell \lambda}_{(q+1)1}$$
$$\left.-i\left[\zeta_{j}\partial_{\eta_{j}}
-\Omega_{j}\partial_{\zeta_{j}}\right] \psi^{\ell \lambda}_{(q-1)0}
-i\left[\zeta_{j}\partial_{\xi_{j}}
+\omega_{j}\partial_{\zeta_{j}}\right] \psi^{\ell \lambda}_{q0}\right\}
=-\left(2/\hbar^{2}\right)\left[E-V\right] \psi^{\ell \lambda}_{q1},
\eqno (11b) $$
$$\bigtriangleup  \psi^{\ell \lambda}_{q \tau}(\xi,\eta,\zeta)=
\left\{4\xi_{1}\partial^{2}_{\xi_{1}}+4\eta_{2}\partial^{2}_{\eta_{2}}
+\left(\xi_{1}+\eta_{2}\right)\partial^{2}_{\xi_{2}}
+4\xi_{2}\left(\partial_{\xi_{1}}
+\partial_{\eta_{2}}\right)\partial_{\xi_{2}}
+6\left(\partial_{\xi_{1}}+\partial_{\eta_{2}}\right)\right.$$
$$+\displaystyle \sum_{j=3}^{N-1}~\left[\xi_{1}\partial^{2}_{\xi_{j}}
+\eta_{2}\partial^{2}_{\eta_{j}}
+\Omega^{-1}_{2}\left(\eta_{j}\Omega_{j}-\xi_{j}\omega_{j}
+\zeta_{j}^{2}\right) \left(\partial^{2}_{\xi_{j}}
+\partial^{2}_{\eta_{j}}\right) \right.$$
$$+\Omega^{-1}_{2}\left(\Omega^{2}_{2}+\Omega^{2}_{j}+\omega_{j}^{2}
+\xi_{1}\zeta_{j}^{2}+\eta_{2}\zeta_{j}^{2}\right)\partial^{2}_{\zeta_{j}}
+4\left(\xi_{j}\partial_{\xi_{j}}+\zeta_{j}\partial_{\zeta_{j}}\right)
\partial_{\xi_{1}}$$
$$\left.\left.+4\left(\eta_{j}\partial_{\eta_{j}}
+\zeta_{j}\partial_{\zeta_{j}}\right)\partial_{\eta_{2}}
+2\left(\eta_{j}\partial_{\xi_{j}}
+\xi_{j}\partial_{\eta_{j}}\right)\partial_{\xi_{2}}
+2\xi_{2}\partial_{\xi_{j}}\partial_{\eta_{j}}\right]
\right\} \psi^{\ell \lambda}_{q \tau}(\xi,\eta,\zeta). \eqno (11c) $$

Due to the limited size of a letter, we have to leave the proof of
the theorem and the detailed calculation elsewhere. When
establishing BF we arbitrarily choose two Jacobi coordinate
vectors ${\bf R}_{1}$ and ${\bf R}_{2}$. Those two vectors may be
replaced with any other two Jacobi vectors. One may change the
choice according to the characteristics of the physical problem
under study, such as some or all particles in the quantum $N$-body
problem are the identical particles.

In deriving the generalized radial equations, the key is to
discover the base-functions 
$Q_{q}^{\ell \tau}({\bf R}_{1},{\bf R}_{2})$ of the angular 
momentum and to choose the right internal variables, some of 
which have odd parity. From Eq. (9) we see that only finite 
number of partial angular momentum states are involved in 
constructing the base-functions 
$Q_{q}^{\ell \tau}({\bf R}_{1},{\bf R}_{2})$. Namely, the 
contributions from the remaining partial angular momentum 
states have been incorporated into those from the generalized 
radial functions.

The two features in this method, that the numbers of both
functions $\psi^{\ell \lambda}_{q\tau}(\xi, \eta, \zeta)$ and
equations are finite, and they depend only on $(3N-6)$ internal
variables, are important for calculating the energy levels and
wavefunctions in a quantum $N$-body problem. In fact, in the
numerical experiment for the quantum three-body problem by the
series expansion, much fewer terms have to be taken to achieve the
same precision of energy than with other methods. The calculation
error will be less in comparison with the method to truncate the
series on the partial angular momentum states. As the number of
the particles in the system increases, we believe, to remove three
independent variables will greatly decrease the calculation
capacity requirement.

\vspace{2mm}
\noindent
{\bf ACKNOWLEDGMENTS}. The authors would like to thank Prof.
Hua-Tung Nieh and Prof. Wu-Yi Hsiang for drawing their attention
to the quantum few-body problems. This work was supported by
the National Natural Science Foundation of China and Grant
No. LWTZ-1298 of the Chinese Academy of Sciences.

\end{document}